\documentclass[aps,pre,floatfix,superscriptaddress,showpacs,nobibnotes,twocolumn]{revtex4-2}
\usepackage{amsmath,amssymb,graphicx,psfrag}

\usepackage{xcolor}
\definecolor{later}{rgb}{0.7, 0.7, 0.7}

\newcommand{\blue}[1] {{ \color{blue} #1}}

\newcommand{\Pe}{\textrm{Pe}}
\usepackage{textcomp}
\newcommand{\angb}[1]{\ifmmode\langle#1\rangle\else\textlangle#1\textrangle\fi}

\addtocounter{secnumdepth}{2}

\makeatletter
\def\p@subsection{}
\def\p@subsubsection{}
\makeatother

\begin{document}

\title{ 
Scaling laws for two-dimensional dendritic crystal growth in a narrow channel
}

\author{Younggil Song}
\affiliation{Department of Physics and Center for Interdisciplinary Research
on Complex Systems, Northeastern University, Boston, MA 02115 USA}
\affiliation{
Materials Science Division, Lawrence Livermore National Laboratory, Livermore, CA 94550, USA
}

\author{Damien Tourret}
\affiliation{IMDEA Materials Institute, Getafe, 28906 Madrid, Spain}

\author{Alain Karma}
\email[Corresponding author: ]{a.karma@northeastern.edu}
\affiliation{Department of Physics and Center for Interdisciplinary Research
on Complex Systems, Northeastern University, Boston, MA 02115 USA}

\date{\today}

\begin{abstract}

We investigate analytically and computationally the dynamics of 2D needle crystal growth from the melt in a narrow channel. 
Our analytical theory predicts that, in the low supersaturation limit, the growth velocity $V$ decreases in time $t$ as a power law $V \sim t^{-2/3}$, which we validate by phase-field and dendritic-needle-network simulations. 
Simulations further reveal that, above a critical channel width $\Lambda \approx 5l_D$, where $l_D$ the diffusion length, needle crystals grow with a constant $V<V_s$, where $V_s$ is the free-growth needle crystal velocity, and approaches $V_s$ in the limit $\Lambda\gg l_D$. 

\end{abstract}

\maketitle


Solid-liquid interfaces exhibit different patterns depending on their growth conditions~\cite{Kurz19}. 
When unconstrained, i.e. growing freely in an infinite domain at a given driving force (e.g. undercooling, supersaturation), a dendritic crystal is known to reach a steady growth velocity. 
The growth velocity and tip morphology of such a free dendrite are given by combining solute and/or heat conservation \cite{Ivantsov47} with 
the solvability condition that incorporates the effect of the anisotropy of the interface excess free energy  of the solid-liquid interface (surface tension)~\cite{Langer87, Langer89, Barbieri89, BAmar93}.
When the dendrite is free of lateral interactions from neighboring crystal, secondary (or higher-order) branches can emerge due to thermal fluctuations~\cite{Pieters86, Barber87, Langer87PRA, Brener93, BrenerTemkin95}, which leads to the formation of complex microstructures. 
In contrast, the constrained growth of a dendrite within a channel reduces to the well-known Saffman-Taylor problem~\cite{KesslerKoplikLevine86, Langer89, BrenerEtAl88, CouderEtal86, Kupferman95, BrenerMuller93, BAarBrener95}.
The resulting crystal shape is a symmetric finger or an unstable widening cell above or below a critical undercooling, respectively~\cite{SabouriGhomi01, brenerKessler02, BrenerMuller93, Kupferman95}. 
The threshold between these morphologies are linked to the surface tension~\cite{brenerKessler02, BrenerMuller93, Kupferman95}. 

Theoretical and numerical investigations of 2D and 3D crystal growth in a channel~\cite{KesslerKoplikLevine86, Langer89, BrenerEtAl88, CouderEtal86, Kupferman95, BrenerMuller93, BAarBrener95, Xing15, KassnerEtAl10} have mostly focused on the dendrite tip morphology for a high undercooling ($\Delta$) or solute supersaturation ($\Omega$). 
Exploring lower undercoolings, two-dimensional phase-field studies~\cite{SabouriGhomi01} have shown that a dendrite in a channel slows down with a power-law of time below an undercooling $\Delta=0.5$, along with a widening of the tip. 
For a higher $\Delta$, the dendrite shows a steady state growth with a symmetric dendrite tip. 
The transition between the two growth regimes is close to the morphological threshold $\Delta=0.49$ of a symmetrical finger~\cite{brenerKessler02, brenerKessler02,Kupferman95}. 

While these studies have shown the existence of these two growth regimes (power-law deceleration and a steady state growth)~\cite{SabouriGhomi01, Xing15}, an analytical theory for the decelerating dynamics and the transition between regimes remains lacking. 
In this letter, we study the growth of a 2D binary alloy dendrite in a channel for a low solute supersaturation $\Omega$ --- readily generalizable to the growth of a pure element at a low undercooling $\Delta$. 
We derive an analytical law for the growth of the confined dendrite by combining the solvability condition~\cite{Langer87,Langer89,Barbieri89,BAmar93} and solute conservation considering a sharp needle-like crystal~\cite{TourretKarma13}. 
Our analytical derivation predicts that the dendrite slows down with time $t$ as $V \sim A t^{-2/3}$, where the prefactor $A$ depends directly and nonlinearly upon the channel width $\Lambda$. 
We verify the theory via two numerical approaches, namely using phase-field (PF) ~\cite{Karma01,Echebarria04} and the dendritic-needle-network (DNN) models ~\cite{TourretKarma13,TourretKarma16}. 


The PF approach is a powerful computational method to simulate complex interface dynamics in various phenomena~\cite{LQChen02,Boettinger02}. 
The method relies on a ``phase field", or order parameter, which has a given value within different phases or grains, and varies continuously through interfaces.
In solidification, taking into account anisotropic properties of solid-liquid interfaces, PF simulations can reproduce the evolution of complex anisotropic structures~\cite{Kobayashi93,Echebarria10}.
An asymptotic analysis permits the use of a diffuse interface much wider than the actual physical interface~\cite{KarmaRappel98}. 
For an alloy, the introduction of an ``anti-trapping'' term corrects the numerically-induced solute trapping due to the artificially wide interface, thus ensuring quantitative and computationally-efficient predictions~\cite{Karma01,Echebarria04}.
Combined with additional numerical techniques, such as parallelization and nonlinear preconditioning \cite{Glasner01}, resulting calculations provide quantitative simulations in two (2D) and three (3D) dimensions, at time and length scales directly comparable to experiments~\cite{Bergeon13, TourretKarma15, Clarke17, Tourret17}. 

The DNN model~\cite{TourretKarma13,TourretKarma16} was developed to simulate dendrite growth at low $\Omega$.
In this regime, PF simulations are challenging, due to the extreme separation of scale between the dendrite tip radius $\rho$ and the solute diffusion length $l_D \equiv D/V$, where $D$ is the diffusivity in the liquid and $V$ is the dendrite tip velocity. 
While PF typically requires a grid spacing much lower than the dendrite tip radius, DNN simulations remain accurate using a grid spacing close to, or even above, a dendrite tip radius~\cite{TourretKarma13, TourretKarma16}, which provides a great computational advantage. 

The DNN model consists of solving the transport of the solute field $u$ in the liquid phase (e.g., the diffusion equation), interacting with a hierarchical network of needle-like branches at equilibrium. 
The growth velocity of each dendritic tip is obtained by combining solute balances at two distinct length scales~\cite{TourretKarma13}.
At a scale $\gg\rho$, curvature effects can be neglected and a dendritic branch can be represented as a sharp line segment at fixed equilibrium concentration $u = 0$. 
On the other hand, at a scale $\ll l_D$, the solute field relaxes fast enough to assume a Laplacian $u$ field in the vicinity of the tip. 
The analytical solution to this problem~\cite{DerridaHakim92} exhibits a square-root singularity of the normal gradient of the field in the vicinity of the tip with a ``flux intensity factor'' $\mathcal F$ that is a measure of the incoming flux towards the dendrite tip. 
Using an analogy with fracture mechanics, namely with the stress field ahead of a crack tip submitted to anti-plane shear (mode III) loading, $\mathcal F$ can be calculated with the J-integral commonly used in fracture mechanics~\cite{Rice68,BAmar02,TourretKarma13}.
The resulting conservation equation reads
\begin{equation}
\label{eqn:FIFcondition}
\rho V^2 = \frac{2 D^2 \mathcal{F}^2 }{d_0} \, ,
\end{equation}
where $d_0$ is the capillary length. 
By combining the above equation and the solvability condition~\cite{Langer87,Langer89,Barbieri89,BAmar93},
\begin{equation}
\label{eqn:Solvcondition}
\rho^2 V = \frac{2 D d_0}{ \sigma } \, ,
\end{equation}
where $\sigma$ is the tip selection parameter fixed by the interface anisotropy~\cite{Barbieri89}, the tip radius and growth velocity of a needle tip is given by~\cite{TourretKarma13}
\begin{align}
\label{eqn:FIF_radius}
\frac{\rho(t)}{d_0} &= \left[ \frac{ 2 }{ \sigma^2 \mathcal{F}(t)^2} \right]^{1/3} \, ,
\\
\label{eqn:FIF_velocity}
\frac{V(t) d_0 }{D} &= \left[ 2 \sigma \mathcal{F}(t)^4 \right]^{1/3} \, .
\end{align}
The transient evolution of the solute field around the dendrite tip is captured by the time-dependent $\mathcal{F}(t)$.

For a needle-like dendrite at $\Omega\ll 1$, the 2D Ivantsov solution~\cite{Ivantsov47} reduces to $\Omega\approx\sqrt{\pi\Pe}$, where $\Pe=\rho V/(2D)$ is the P\'eclet number.
Then, the steady radius $\rho_s$ and velocity $V_s$ of a free dendrite become~\cite{TourretKarma13}
\begin{align}
\label{eqn:Steady_radius}
\frac{\rho_s}{d_0} &= \frac{ \pi }{ \sigma \Omega^2} \,, 
\\
\label{eqn:Steady_velocity}
\frac{ V_s d_0}{D } &= \frac{ 2 \sigma \Omega^4}{ \pi^2 } \, .
\end{align}


Let a needle-like dendrite grow within a channel of width $\Lambda$. 
For $\Lambda \ll l_D$, the solute field surrounding the needle tip is essentially Laplacian with $\nabla^2 u=0$. 
Applying the J-integral around the tip in a channel yields that $\mathcal{F}(t)$ depends on the solute gradient, $\partial u/\partial x$, ahead of the tip in the needle growth direction $x$ as~\cite{Rice68,BAmar02,TourretKarma13}
\begin{equation}
\label{eqn:FIFsq}
\mathcal{F}(t)^2 = \frac{ \Lambda d_0}{ 2 \pi } \left( \frac{ \partial u }{ \partial x}\right)^2 \, .
\end{equation}
For $\Lambda \ll l_D$, the solute concentration profile can also be considered one-dimensional (1D), such that the gradient in front of the needle follows the Zener approximation~\cite{Zener49, Ivantsov47}
\begin{equation}
\frac{ \partial u }{ \partial x}  \approx \frac{ \Omega }{ \sqrt{\pi D t} } \, .
\end{equation} 
Combining these last two equations with Eq.~\eqref{eqn:FIF_velocity}, the needle tip velocity can be expressed as 
\begin{align}
\label{eqn:Vorigin}
V(t) = \left ( \frac{ \sigma \Lambda^2 }{2 \pi^4 } \frac{ D }{ d_0 }  \Omega^4 \right)^{1/3} t^{-2/3}  \, .
\end{align}
Normalizing time and space using theoretical steady-state values of $\rho_s$ (Eq.~\eqref{eqn:Steady_radius}) and $V_s$ (Eq.~\eqref{eqn:Steady_velocity}) leads to
\begin{align}
\label{eqn:Vnarrow}
\widetilde{V} (\tilde{t}) =  \frac{V(t)}{V_s} = \left( \frac{\widetilde{\Lambda}}{2 \pi \tilde{t} } \right)^{2/3}  \, ,
\end{align}
where $\tilde{t} = t V_s/\rho_s$ and $\widetilde{\Lambda} = \Lambda/\rho_s$.


\begin{figure}[!b]
\includegraphics[width=3.4 in]{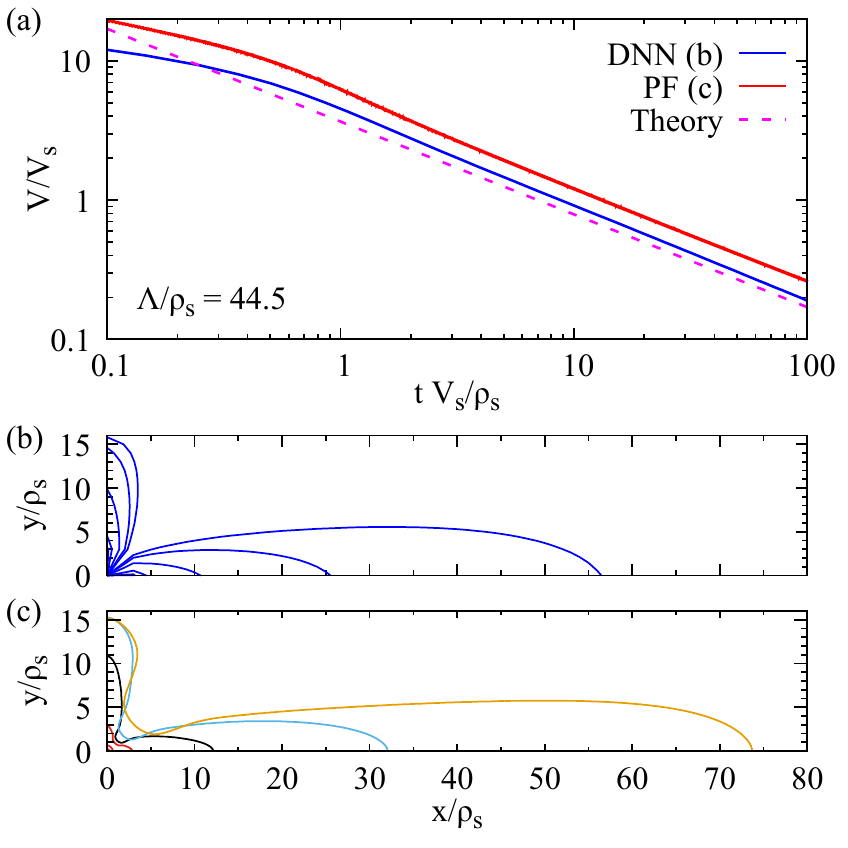}
\caption{
Calculated growth velocities (a) and morphological evolution (b)-(c) of a dendrite in a channel of width $\widetilde \Lambda=44.5$ at a solute supersaturation $\Omega=0.05$.
Both DNN (blue solid line) and PF (red solid line) results exhibit a power law close to the analytical theory, Eq.~\eqref{eqn:Vnarrow} (pink dashed line).
Morphologies in (b) and (c) show solid-liquid interface locations at $\tilde{t}=0.01$, 0.1, 1, 10, and 100 from the DNN and PF simulations, respectively. 
In panel (c), different colors denote different times $\tilde{t} <$ 1 (red), $\tilde{t}=$ 1 (black), 10 (cyan), and 100 (yellow).
}
\label{fig:Narrow}
\end{figure}

In order to validate the scaling law \eqref{eqn:Vnarrow} for a dendrite growing in a narrow channel at low $\Omega$, we first performed both PF and DNN simulations at $\Omega=0.05$, whose results for $\widetilde{\Lambda} = 44.5$ are shown in Fig.~\ref{fig:Narrow}(a). 
Both PF (red solid line) and DNN (blue solid line) predictions follow Eq.~\eqref{eqn:Vnarrow} (pink dashed line). 
While the DNN needle effectively remains a sharp line interacting with the solute field, its virtual thickness can still be estimated by time integration of the fluxes along its sides (see details in Supplementary Material~\cite{SuppleM} and Ref.~\cite{TourretKarma13}). 
The resulting interface and its time evolution, shown at $\tilde{t} =  0.01, 0.1, 1, 10,$ and $100$ in Fig.~\ref{fig:Narrow}(b), agrees reasonably with the interface predicted by the PF method (Fig.~\ref{fig:Narrow}(c)).

\begin{figure}[!b]
\includegraphics[width=3.4 in]{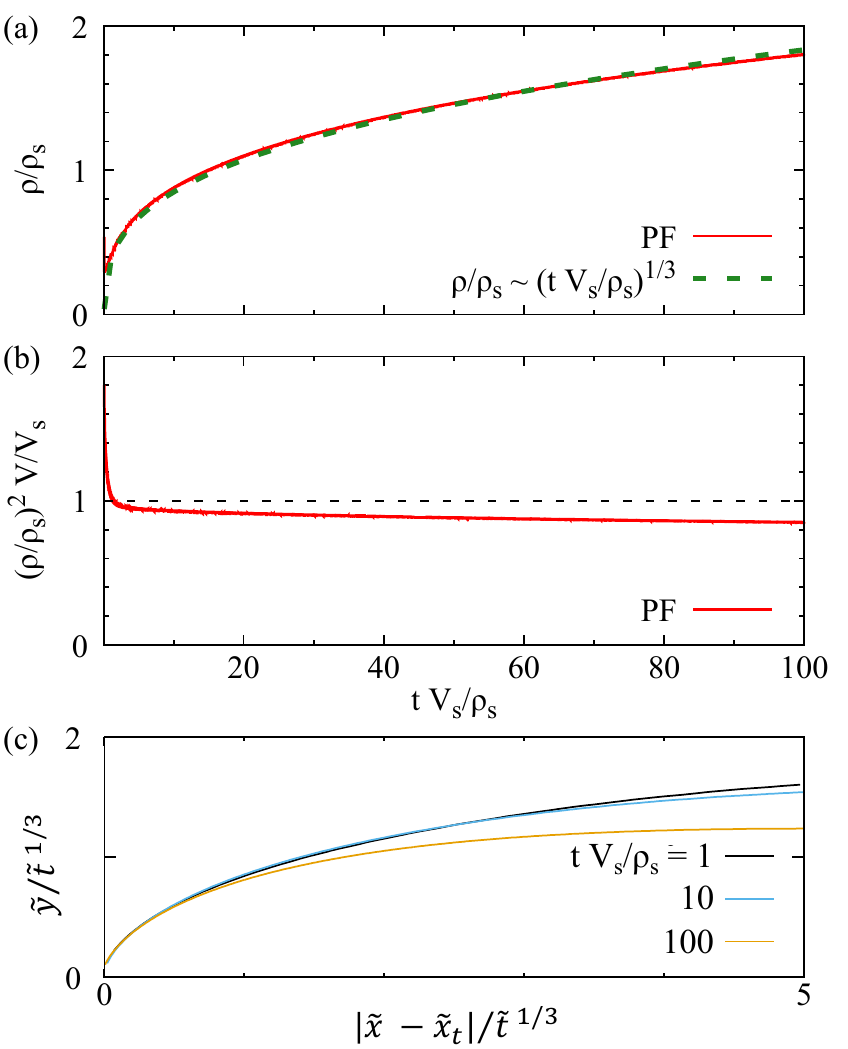}
\caption{
Time-dependent tip radius $\tilde{\rho}$ (a) and $\tilde{\rho}^2 \widetilde{V}$ (b) of a dendrite in $\widetilde \Lambda=44.5$ at $\Omega=0.05$. 
Panel (a) shows $\tilde{\rho}$ from the PF simulation of Fig.~\ref{fig:Narrow} (red line) compared to a power law~$\sim \tilde{t}^{1/3}$ (green dashed line). 
In (b), $\tilde{\rho}^2\widetilde{V}$ of the PF dendrite within a narrow channel (red line) is compared to the expected value of 1 (thin dashed line) for a free dendrite. 
Panel (c) compares the scaled tip morphologies $\tilde{y}/\tilde{t}^{1/3}$ against $|\tilde{x} - \tilde{x}_t| /\tilde{t}^{1/3}$ at three different times $\tilde{t}=$ 1 (black), 10 (cyan), and 100 (yellow). 
}
\label{fig:PFNarrow}
\end{figure}

Because the DNN dendrite remains infinitely sharp, thus matching the underlying assumption of Eq.~\eqref{eqn:Vnarrow}, the DNN-predicted velocity has a prefactor closer to that of Eq.~\eqref{eqn:Vnarrow} than that from the PF simulation.
In the latter, the dendrite actually thickens, thus partially filling the channel and effectively increasing the solute gradient in the $x$ direction and yielding a higher velocity than that predicted for a sharp needle.
These results are consistent with previous PF studies~\cite{SabouriGhomi01, Xing15}, which showed that the power law deceleration observed at a relatively low P\'eclet number disappeared when the undercooling was increased and the dendrite was progressively filling up the channel.

In the PF simulation of Fig.~\ref{fig:Narrow}(c), the evolution of the dendrite tip radius and its morphology are also relevant to the filling of the channel. 
As shown in Fig.~\ref{fig:PFNarrow}(a), the tip radius of the confined dendrite (red solid line) increases as $\tilde{\rho} \sim \tilde{t}^{1/3}$ (green dashed line), which is consistent with the constancy of $\rho^2 V$ predicted by solvability theory~\cite{Langer87,Langer89,Barbieri89,BAmar93}. 
Indeed, the calculated evolution of $\rho^2 V$, plotted as red line in Fig.~\ref{fig:PFNarrow}(b), quickly approaches a constant value slightly lower but close to $\tilde{\rho}^2 \widetilde{V} = 1$ (black dashed line), i.e., close to the free dendrite steady state with $\rho \approx \rho_s$ and $V \approx V_s$ \cite{PlappKarma, TourretKarma13}. 
For the confined dendrite, $\tilde{\rho}^2 \widetilde{V}$ appears to decrease slowly soon after it reaches $\tilde{\rho}^2 \widetilde{V} \approx 1$, but remains close to 1 during the entire simulation. 

Since the tip velocity evolves as $\widetilde V \sim \tilde t^{-2/3}$, the tip position $\tilde x_t$ increases as $ \tilde{x}_t\sim \tilde t^{1/3}$, i.e., with a similar power law as the tip radius, $\tilde{\rho} \sim \tilde{t}^{1/3}$.
One can thus assess whether the dendrite tip retains a self-similar shape over time by plotting $|\tilde{x} - \tilde{x}_t|  /\tilde{t}^{1/3}$ against $\tilde{y}/\tilde{t}^{1/3}$, as done in Fig.~\ref{fig:PFNarrow}(c).
There, the scaled dendrite shapes at different times $\tilde{t}=$ 1 (black), 10 (cyan), and 100 (yellow) show that the immediate vicinity of the tip remains almost unchanged.
However, later in the simulation, as the dendrite starts to fill the channel (e.g., $\tilde{t}=100$), the effect of confinement becomes apparent with a relatively narrower shape of the dendrite further from the tip. 

\begin{figure}[!h]
\includegraphics[width=3.4 in]{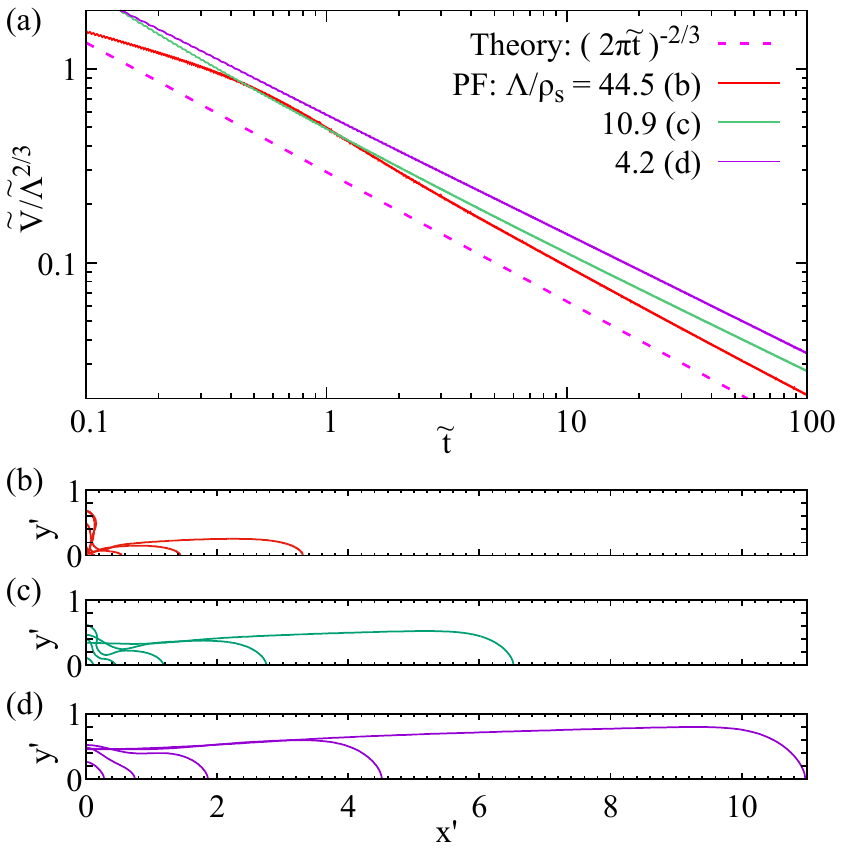}
\caption{
PF-predicted growth velocities (a) and morphological evolutions (b-d) of dendrites for different channel widths $\widetilde \Lambda=44.5$ (red), $10.9$ (green), and $4.2$ (purple solid line), compared to the analytical law with $\widetilde V\sim \tilde{t}^{-2/3}$ (pink dashed line).
Panels (b), (c), and (d) show snapshots of the interface locations at $\tilde t=0.01$, 0.1, 1, 10, 100 for $\widetilde \Lambda=44.5$, 10.9 and 4.2, respectively. 
They are scaled with respect to $\Lambda/2$, i.e., with $x' = 2 x / \Lambda$ and $y' = 2 y / \Lambda$. 
}
\label{fig:TooNarrow}
\end{figure}

In order to investigate the effect of the dendrite volume (area) on its growth dynamics, we performed additional PF simulations with channel widths $\widetilde{\Lambda} = 4.2$ and $10.9$. 
In Fig.~\ref{fig:TooNarrow}(a), $\widetilde{V}$ is scaled by $\widetilde{\Lambda}^{2/3}$, in order to represent all simulations on the same graph, comparing them with the expected $\widetilde{V}/\widetilde{\Lambda}^{2/3} = (2 \pi \tilde{t} )^{-2/3}$ (pink dashed line) from Eq.~\eqref{eqn:Vnarrow}. 
%
%
Solid lines show results with $\widetilde{\Lambda} \simeq$ 44.5 (red), 10.9 (green), and 4.2 (purple). 
A wider channel results in a longer initial transient time.
Nevertheless, once a power-law regime is reached ($\tilde{t}\geq1$) the widest channel leads to the best agreement with Eq.~\eqref{eqn:Vnarrow}.

Interface shapes at $\tilde{t} = 0.01$, 0.1, 1, 10, and $100$ are drawn in Fig.~\ref{fig:TooNarrow}(b)-(d) for $\widetilde{\Lambda} = 44.5$, 10.9, and 4.2, respectively, with space coordinates normalized by the channel half-width.
As expected, narrower channels (Fig.~\ref{fig:TooNarrow}(d)) get filled by the dendrite faster than wider channels. 
These results confirm that, for a given $\Omega$, the dendrite velocity deviates from the proposed growth law when $\Lambda$ decreases, as the assumption that $\Lambda\gg\rho$ is no longer valid.

Regardless of the channel width, dendrites slow down with an exponent close to the predicted $V \sim t^{-2/3}$ law.
Once the effect of confinement becomes predominant, as long as $\rho\ll\Lambda$, the constancy of $\rho^2 V$~\cite{Langer87,Langer89,Barbieri89,BAmar93} implies that $\rho$ increases with a corresponding $\rho \sim t^{1/3}$ as shown in Fig.~\ref{fig:PFNarrow}(a).
Hence, as $\rho$ increases, the problem approaches that of a 1D planar interface, known to follow $V\sim t^{-1/2}$~\cite{Zener49, Pelce88,BrenerMuller93}. 
Hence, even though dendrites in Fig.~\ref{fig:TooNarrow} are expected to retain a finger morphology rather than asymptotically tending toward a planar interface, the higher exponent exhibited by the thicker dendrites (i.e. narrower channel, e.g. Fig.~\ref{fig:TooNarrow}(d)) likely indicates an intermediate state between dendritic ($V\sim t^{-2/3}$) and planar ($V\sim t^{-1/2}$) growth.

\begin{figure*}[!ht]
\includegraphics[width=6.8 in]{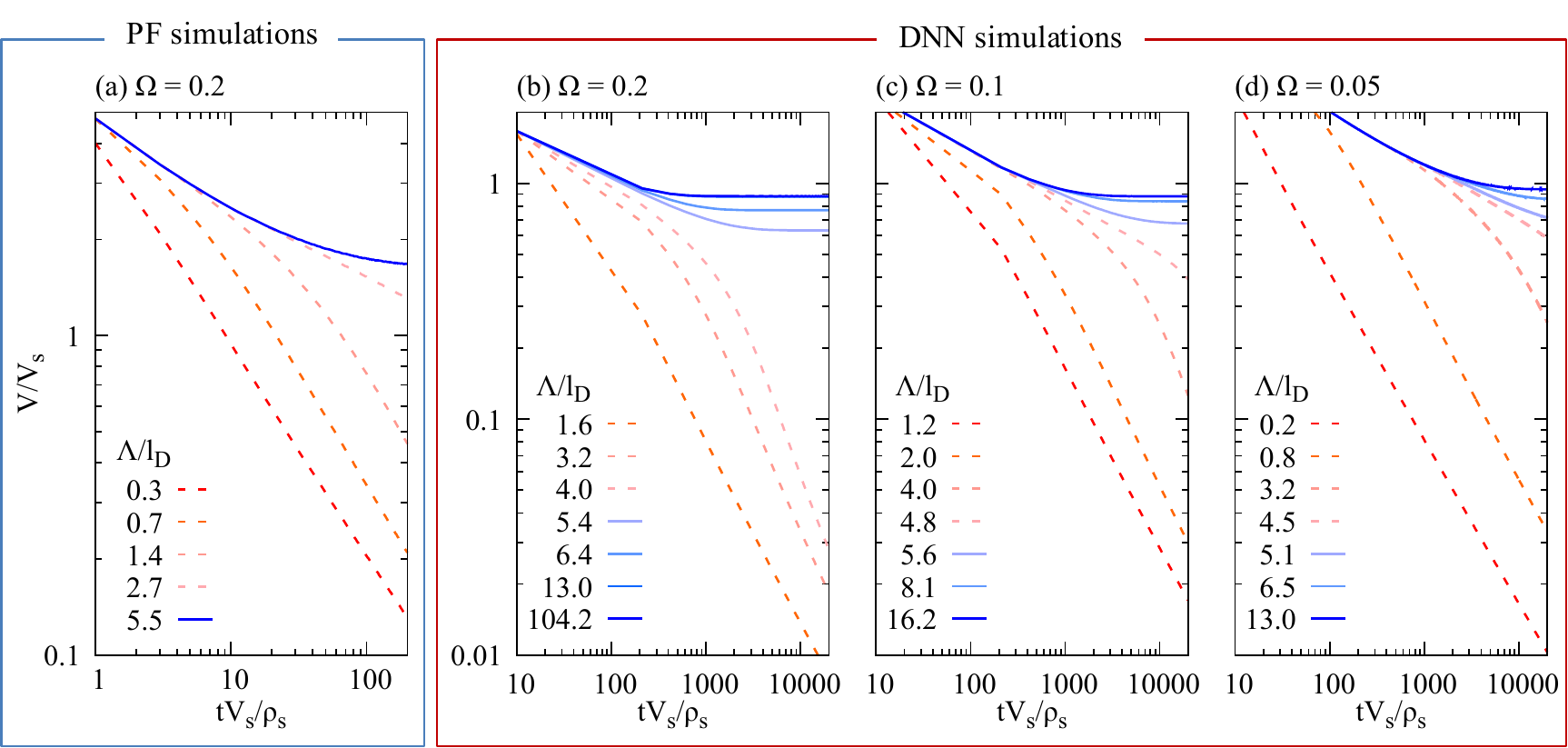}
\caption{
Growth velocities of a dendrite within different channel widths $\Lambda$. 
(a) shows PF results at $\Omega=0.2$; (b), (c), and (d) show DNN results at $\Omega=0.2$, 0.1, and 0.05, respectively. 
The channel widths are scaled with respect to the diffusion length ${l}_D / \rho_s = 39.3$, 157.1, and 628.3 for $\Omega=0.2$, 0.1, and 0.05, respectively.
Within a narrow channel, the dendrite is decelerated with the power law (red dashed lines). Otherwise, it approaches a steady state (blue solid lines).
}
\label{fig:TooWide}
\end{figure*}

If the channel is very wide, the resulting free dendrite should not be affected by the domain boundaries and thus reach a steady state velocity $\widetilde{V}=1$. 
Consequently, for a given supersaturation, we expect that there exists a critical channel width $\Lambda_c$ that delimits these two growth regimes.
We performed simulations with various $\Lambda$ at a given $\Omega$ in order to identify ${\Lambda}_c$, namely with $\Omega=0.2$ (PF) and $\Omega = 0.2$, 0.1, and 0.05 (DNN). 
The diffusive interaction between the dendrite and the channel boundaries is expected to be the key mechanism that determines whether the dendrite will decelerate or reach a steady state. 
Hence, results presented in Fig.~\ref{fig:TooWide} are scaled with respect to the diffusion length ${l}_D$. 
For $\Omega=0.2$, 0.1, and 0.05, the diffusion lengths of a free dendrite are ${l}_D / \rho_s = 39.3$, 157.1, and 628.3, respectively. 

In PF simulations (Fig.~\ref{fig:TooWide}(a)), growth velocities in the three narrowest channels, namely $\Lambda/l_D=$ 0.3, 0.7 and 1.4 (red dashed lines) decelerate towards the proposed power law. 
As observed earlier, as ${\Lambda}$ is wider, it takes a longer time to reach the $\tilde{t}^{-2/3}$ growth regime.  
In particular, at ${\Lambda}/l_D = 2.7$ (lightest red dashed line), the dendrite is still well within a transition regime towards the power law at the end of the simulation.
On the other hand, the dendrite velocity for ${\Lambda}/l_D = 5.5$ (blue solid line) does not show any inflexion during the simulated time, and seems to approach a steady-state velocity. 

DNN simulations allow us to explore a broader range of parameters, and show the similar results, namely with $\Omega=0.2$, 0.1, and 0.05. 
Results, shown in Fig.~\ref{fig:TooWide}(b)-(d), exhibit the same behavior as in PF simulations (Fig.~\ref{fig:TooWide}(a)).
Dendrites in narrower channels (red dashed lines) slow down towards the theoretical power law.
Meanwhile, dendrites in wider channels (blue solid lines) reach a constant growth velocity. 
At $\Omega=0.05$, the distinction between decelerating and steady regime is not obvious within the time range represented in Fig.~\ref{fig:TooWide}(d), but it appears unambiguously looking at longer simulations~\cite{SuppleM}.

Both PF and DNN results exhibit two distinct growth behaviors: the power law deceleration at low $\Lambda$ and a steady state growth at higher $\Lambda$. 
The crossover between the two distinct growth dynamics occurs at $ 2.7 < \Lambda_c/l_D < 5.5$ in PF simulations and at
$4.8 < \Lambda_c/l_D < 5.1$ according to DNN results. 
We conclude that the critical channel width $\Lambda_c$ is approximately $5\, {l}_D$. 
Interestingly, this crossover at $\Lambda\approx5\, {l}_D$ is consistent with other results showing that the transition between quasi-2D confined growth and 3D growth regimes occurs at a sample height of approximately $5\, {l}_D$ in 3D DNN simulations~\cite{DNNCET}. 

One unexpected observation from Fig.~\ref{fig:TooWide} is that, at ${\Lambda}>{\Lambda}_c$, dendrites can reach a steady state velocity that is notably smaller than $V_s$.
In such cases, the channel boundaries still interact with the dendrite even if ${\Lambda}$ is wide enough for a dendrite to reach a steady state. 
In our simulations, the reached steady state velocity are as low as about $0.6 V_s$. 
The channel width necessary for the dendrite to stabilize at a steady velocity close to the free dendrite theoretical velocity $V_s$ can be as high as over a hundred times the diffusion length at relatively high $\Omega=0.2$, as seen in Fig.~\ref{fig:TooWide}(b).
As a dendrite approaches a steady state, its tip morphology is also stabilized. 
The dendrite morphology can thus be described using the solvability condition in Eq.~\eqref{eqn:Solvcondition}~\cite{Langer87,Langer89,Barbieri89,BAmar93} or other approaches~\cite{Almgren93,Brener93,BrenerTemkin95,AlexandrovEtAl}. 


In summary, we developed an analytical theory for transient dendrite growth in a channel that predicts the power law deceleration $V \sim t^{-2/3}$. This  result adds to our theoretical understanding of transient dendrite growth, which has been so far limited to early stage growth in an unconstrained geometry (described by the scaling law $V \sim t^{-2/5}$) \cite{Almgren93}. Both PF and DNN simulations of dendritic growth within a narrow channel show that this power law holds as long as the channel width $\Lambda$ is larger than the dendrite tip radius $\rho$.
PF simulations show that, as the dendrite slows down, following solvability theory~\cite{Langer87,Langer89,Barbieri89,BAmar93}, its radius increases, the filling of the channel by the dendrite increases, and the velocity exponent rises above $-2/3$, thus getting closer to that expected for a planar interface ($V \sim t^{-1/2}$).
On the other hand, when $\Lambda$ is larger than a critical width $\Lambda_c \approx 5 l_D$, dendrites can reach a steady state with a constant velocity.
When $\Lambda$ is slightly higher than $\Lambda_c$, the dendrite can stabilize at a velocity lower than $V_s$. 
This velocity increases towards $V_s$ as $\Lambda$ increases.

Results of this work may be related to the coarsening dynamics of parallel dendritic arms, e.g., after morphological destabilization of a planar interface or among dendritic sidebranches. In such case, an effective channel width is imposed by the presence of neighbor branches, and the growth kinetics of eliminated (i.e., slowing down) branches should follow the $V\sim t^{-2/3}$ power law identified here, until enough branches are eliminated and the spacing among active branches reaches the identified threshold of $\approx 5 l_D$. Ongoing work on extending the present study to the coarsening dynamics of dendritic sidebranches, and resulting scaling laws for grain envelope shapes, will be presented elsewhere.

This study remains to be extended to three dimensions.
PF studies revealed that 3D growth in a narrow channel also exhibits a power law deceleration~\cite{Xing15} with a gradual transition toward a steady growth velocity~\cite{Xing15, DNNCET}. 
The lack of known analytical solution for the diffusion field surrounding an infinitely sharp needle --- similar to that used in 2D in this Letter --- makes the current approach not readily applicable, such that an analytical description of the deceleration regime and the corresponding critical width in 3D remains missing.

\section*{ ACKNOWLEDGEMENT }

This research was primarily supported by NASA grants NNX16AB54G and 80NSSC19K0135.
D.T. acknowledges financial support from the Spanish Ministry of Science through the Ramon y Cajal grant RYC2019-028233-I.
Y.S. acknowledges that this work was partially performed under the auspices of the U.S. Department of Energy by Lawrence Livermore National Laboratory under Contract DE-AC52-07NA27344.



\end{document}


\title{ 
Supplemental Material: \\
Scaling laws for two-dimensional dendritic crystal growth in a narrow channel
}

\author{Y. Song}
\affiliation{Department of Physics and Center for Interdisciplinary Research
on Complex Systems, Northeastern University, Boston, MA 02115 USA}
\affiliation{
Materials Science Division, Lawrence Livermore National Laboratory, Livermore, CA 94550, USA
}

\author{D. Tourret}
\affiliation{IMDEA Materials Institute, Getafe, 28906 Madrid, Spain}

\author{A. Karma}
\email[Corresponding author: ]{a.karma@northeastern.edu}
\affiliation{Department of Physics and Center for Interdisciplinary Research
on Complex Systems, Northeastern University, Boston, MA 02115 USA}

\date{\today}

\maketitle


\section{Simulation models} 

We use two simulation techniques, namely phase-field (PF) and dendritic-needle-network (DNN) models, to simulate a dendrite growth in a channel.

\subsection{Phase-field model } 

The phase-field (PF) method for solidification has yielded quantitative comparison to experiments in two (e.g.,~\cite{GonzalezCincaEtAl05,Echebarria10,TourretKarma15,Tourret17, Guo17}) and three dimensions (e.g.,~\cite{Bergeon13, Tourret15, Clarke17,Tourret17}). 
Here, we use the PF model of a dilute binary alloy within the thin interface limit~\cite{KarmaRappel98}, with a corrective anti-trapping solute current at the interface~\cite{Karma01,Echebarria04}. 
We use a preconditioned phase-field $\psi$ with $\varphi\equiv\tanh(\psi/\sqrt{2})$~\cite{Glasner01,TourretKarma15}, in place of the classical $\varphi$ ($\varphi = +1$ in the solid and -1 in the liquid).
In the PF model, the dimensionless supersaturation $U$ is defined as
\begin{equation}
\label{eqn:PF:U}
U \equiv \frac{1}{1-k} \left[ \frac{ 2 c/c_l^0}{ ( 1+k ) - (1 - k) \varphi } - 1 \right] \, , 
\end{equation}
where $c$ is a solute concentration, and
$k = c_s^0/c_l^0$ is the solute partition coefficient with the equilibrium solute concentrations on the solid $c_s^0$ and on the liquid side $c_l^0$ of a planar interface, respectively. 
After space and time are scaled with respect to the diffuse interface width $W$ and relaxation time $\tau_0$ at the temperature $T_0$~\cite{Echebarria04}, the time evolution of $\psi$ and $U$ follows 
\begin{align}
\label{eqn:phi}
a_s(\theta)^2 \frac{\partial\psi}{\partial t} &= 
\vec\nabla \left[a_s(\theta)^2\right] \vec\nabla\psi  \nonumber \\
&+~ a_s(\theta)^2 \left[  \nabla^2\psi -\varphi\sqrt{2}|\vec\nabla\psi|^2  \right]  \nonumber\\
&+~\sum_{m=x,y}\left[ \partial_m\left( |\vec\nabla\psi|^2 a_s(\theta) \frac{\partial a_s(\theta)}{\partial(\partial_m\psi)} \right) \right] \nonumber\\
&+~ \sqrt 2 \left( \varphi  - \lambda_c (1-\varphi^2) U \right) \, ,
\end{align}
\begin{align}
\label{eqn:U}
&\Big( 1+k-(1-k)\varphi \Big) \frac{\partial U}{\partial t} = 
{D}' \; \vec\nabla\cdot\left[ (1-\varphi)\vec\nabla U\right]  \nonumber\\
&\qquad +~\vec\nabla\cdot\left[ \Big(1+(1-k)U\Big) \frac{(1-\varphi^2)}{2} \frac{\partial\psi}{\partial t} \frac{\vec\nabla\varphi}{|\vec\nabla\varphi|} \right] \nonumber\\
&\qquad +~\Big[ 1+(1-k)U \Big]  \frac{ \partial \varphi}{\partial t} \, , 
\end{align}
with the dimensionless diffusivity 
\begin{align}
\label{eqn:diffusivity}
D' &= D \frac{\tau_0}{W} = a_1 a_2 \frac{W}{d_0} \, ,
\end{align}
and the coupling factor 
\begin{align}
\label{eqn:cfactor}
\lambda_c &= a_1 \frac{W}{d_0} \, .
\end{align}
The constants $a_1=5\sqrt{2}/8$ and $a_2=47/75$ are obtained from the thin interface limit with vanishing kinetic coefficient~\cite{KarmaRappel98}. 
A 2D fourfold anisotropy is imposed on the diffuse interface width and on the phase field relaxation time, using the standard form $a_s(\theta)=1+\varepsilon_4 \cos (4\theta)$, where $\varepsilon_4$ is the surface tension anisotropy strength, and $\theta$ is an angle between the interface normal and a fixed $x$ axis linked to the crystal orientation~\cite{Echebarria04, Echebarria10, TourretKarma15}. 
Equations are solved numerically using finite differences on a grid size $\Delta x$ and an explicit time stepping $\Delta t$.
Further details about the model were published elsewhere~\cite{Echebarria04,TourretKarma15}.

\subsection{Dendrite-needle-network model } 

The DNN model consists in solving the transport of the solute field $u$ in the liquid phase, e.g., the diffusion equation, interacting with a network of needle-like branches at equilibrium. 
The growth velocity of each dendritic tip is obtained by combining solute balances at two distinct length scales, i.e., using two relations~\cite{TourretKarma13} summarized below.

The DNN formulation used here considers infinitely sharp needle-like branches~\cite{TourretKarma13}.
At a low supersaturation $\Omega$, since $l_D\gg\rho$, one can write a solute balance at an intermediate scale between $\rho$ and $l_D$. 
At a scale $\gg\rho$, curvature effects can be neglected and a dendritic branch can be represented as a sharp line segment at fixed equilibrium concentration $u = 0$. 
On the other hand, at a scale $\ll l_D$, the solute field relaxes fast enough to assume a Laplacian field $u$ in the vicinity of the tip. 
The analytical solution to this problem~\cite{DerridaHakim92} exhibits a square-root singularity of the normal gradient of the field in the vicinity of the tip as
\begin{equation} 
\left. \frac{\partial u}{\partial y} \right|_{y=0} = \frac{\mathcal F}{\sqrt{d_0 (x_t-x)}} 
\end{equation}
when the position $x$ tends to the tip position $x_t$. 
The flux intensity factor $\mathcal F$ is a measure of the incoming flux toward the tip, defined as
\begin{equation}
\label{eqn:FIF_basic}
\mathcal{F} \equiv \lim_{x \to x_t} \left. \sqrt{d_0(x_t - x)} \, \frac{\partial u}{ \partial y } \right|_{y=0} \, .
\end{equation}
In 2D, $\mathcal F$ can be directly calculated with an integral along a contour $\Sigma$ around the tip similar to the J-integral commonly used in fracture mechanics~\cite{Rice68,BAmar02,TourretKarma13}  	
\begin{equation} 
\label{eqn:FIFcontour}
{\mathcal F}^2 = \frac{d_0}{2 \pi} \int_{\Sigma} \Big[ \Big ( (\partial_{{x}}u)^2 - (\partial_{{y}}u)^2 \Big )n_{{x}} 
+ 2\ \partial_{{x}}u\partial_{{y}}u\, n_{{y}} \Big] \mathrm{d} \Sigma \, .
\end{equation}
Further away from the tip, the interface position $y_i(t)$ is assumed to follow the solution of a 1D diffusion problem in the direction normal to the branch. 
Using the flux along the sharp needle ($y=0$) as an approximation for the flux at the interface ($y=y_i(t)$), we write
\begin{equation} 
\frac{\mathrm{d}y_i(t)}{\mathrm{d}t} \approx D\, \left.\frac{\partial u}{\partial y} \right|_{y=0} = D \frac{\mathcal F }{ \sqrt{d_0 (x_t-x)}} \, .
\end{equation}
Then, assuming that, at this scale with $x_t-x \ll D/V$, the tip moves in a quasi-steady regime, we can use the change of variable $x_t-x=Vt$, and obtain the relation linking $\rho$ and $V$~\cite{TourretKarma13}
\begin{equation}
\label{eqn:FIFcondition}
\rho V^2 = \frac{2 D^2 \mathcal{F}^2}{d_0} \, .
\end{equation}
This relation, obtained here for an infinitely sharp needle using an analytical solution to the Laplace equation and an analogy to fracture mechanics, may also be obtained for a parabolic tip shape using solely solute mass conservation considerations~\cite{TourretKarma16,Tourret19}.

The second relation, established at the scale of the dendritic tip radius $\rho$, is the well-established solvability condition~\cite{Langer87,Langer89,Barbieri89,BAmar93},
\begin{equation}
\label{eqn:Solvcondition}
\rho^2 V = \frac{2 D d_0}{ \sigma } \, ,
\end{equation}
where $\sigma$ is the tip selection parameter fixed by the interface anisotropy~\cite{Barbieri89}. 
This relation for the existence of a steady growing solution of a parabolic tip has been validated by PF simulations, which have shown that $\sigma$ reaches a constant value as soon as parabolic tips start emerging at the very early stage of dendritic growth~\cite{PlappKarma00}. 
Therefore, in the DNN model, Eq.~\eqref{eqn:Solvcondition} is used during both steady state and transient growth stages.

Combining Eqs.~\eqref{eqn:FIFcondition} and~\eqref{eqn:Solvcondition}, the growth dynamics of each needle tip is given by
\begin{align}
\label{eqn:FIF_radius}
\frac{\rho(t)}{d_0} &= \left( \frac{ 2 }{ \sigma^2 \mathcal{F}(t)^2} \right)^{1/3} \, ,
\\
\label{eqn:FIF_velocity}
\frac{V(t) d_0 }{D} &= \left( 2 \sigma \mathcal{F}(t)^4 \right)^{1/3} \, ,
\end{align}
where the transient evolution of the solute field surrounding the dendrite tip is captured by the time-dependent flux intensity factor $\mathcal{F}(t)$.

For the sake of numerics and generality, we scale time and space with respect to a theoretical stationary radius $\rho_s$ and velocity $V_s$ of a single free dendrite tip.
For a supersaturation $\Omega\ll 1$ and an infinitely thin needle-like dendrite, the two-dimensional Ivantsov solution~\cite{KurzFisher, Ivantsov47} may be approximated as $\Omega\approx\sqrt{\pi\Pe}$.
Combining the latter with solvability Eq.~\eqref{eqn:Solvcondition}, we obtain 
\begin{align}
\label{eqn:Steady_radius}
\frac{\rho_s}{d_0} &= \frac{ \pi }{ \sigma \Omega^2} \,, 
\\
\label{eqn:Steady_velocity}
\frac{ V_s d_0}{D } &= \frac{ 2 \sigma \Omega^4}{ \pi^2 } \, .
\end{align}
The resulting dimensionless tip radius $\tilde{\rho}\equiv\rho/\rho_s$ and velocity $\widetilde{V}\equiv V/V_s$ evolve with time $\tilde{t}\equiv t V_s/\rho_s$ as
\begin{align}
\label{eqn:scaledR}
\tilde{\rho}&=(2 \widetilde{D}^2 \widetilde{\mathcal{F}}^2 )^{-1/3} \, ,
\\
\label{eqn:scaledV}
\widetilde{V}&=(2 \widetilde{D}^2 \widetilde{\mathcal{F}}^2 )^{2/3} \, ,
\end{align}
where $\widetilde{D}=\pi/(2\Omega^2)$, and the scaled flux intensity factor $\widetilde{\mathcal{F}}$, following Eq.~\eqref{eqn:FIFcontour}, is 
\begin{equation} 
\label{eqn:FIFscaledContour}
{\widetilde{\mathcal F}}^2 = \frac{1}{2 \pi}\int_{\Sigma} \Big[ \Big ( (\partial_{{\tilde x}}u)^2 - (\partial_{{\tilde y}}u)^2 \Big )n_{{\tilde x}} + 2\ \partial_{{\tilde x}}u\partial_{{\tilde y}}u\, n_{{\tilde y}} \Big] \mathrm{d} \Sigma .
\end{equation}
Equations \eqref{eqn:scaledR}-\eqref{eqn:FIFscaledContour}, together with the diffusion equation 
\begin{equation} 
\label{eqn:scaledDiff}
\partial_{\tilde t}u = \widetilde D\nabla^2 u
\end{equation}
and the boundary conditions ($u=0$ on each needle and $u=\Omega$ in the liquid far away from the solid) constitute the summary of the 2D DNN model for isothermal growth.

DNN equations are solved numerically using finite differences on a grid size $\Delta x$ with an explicit time step $\Delta t$.
In the ``sharp'' DNN model used here, the needles do not thicken and remain sharp line segments.
Nonetheless, we can estimate the volume of a needle-like branch by integrating over time $\partial u/\partial y |_{y=0}$ along its sides. 
In discrete form, let a needle grow in the $x$ direction, with its tip located at $(i+r,j)$, where $i$ and $j$ are integer grid coordinates and $0 \le r < 1$ describes the progression of the tip position between successive grid points.
At a position $(i',j)$ along the needle length, with $0 < i' \ (= i+r) <i $, the incoming solute flux on the $y+$ side of the dendrite is 
\begin{equation} 
\label{eqn:income}
\vec{I} = - D \frac{ u(i',j) - u(i',j+1)}{ \Delta x } \, ,
\end{equation}
where $u(i',j)=0$ on the needle. 
Moreover, the incoming solute at the tip in the $x$ direction is 
\begin{equation} 
\vec{I}_{t} = D \frac{ u(i+1,j)}{ (1-r) \Delta x } \, .
\end{equation}
Thus, during a time step $\Delta t$, the dendrite volume increases by $\vec{I} \, \Delta t \Delta x$ on each side and $\vec{I}_{t} \Delta t  \Delta x$ at the tip.
The total volume of a dendrite is calculated by integrating these fluxes over time.


\section{Method}

We perform PF and DNN simulations using parameters listed in Table~\ref{tab:parameters}. 
We keep the same values for $k$, $\varepsilon_4$, and $\sigma$ in all simulations. 

\begin{table}[b!]
\caption{ Parameters used for PF and DNN simulations.}
\begin{center}
\begin{tabular}{ c r r }
\hline
\hline
Symbols & PF model  & DNN model
\\ 
\hline
$\Omega$ & 0.05, 0.2 & 0.05, 0.1, 0.2
\\
$k$ & 0.15 & 
\\
$\varepsilon_4$ & 0.03 &  
\\
$\sigma$ & 0.22 & 0.22
\\
$\Delta x$ & 1~($W$)=30, 400~($d_0$) & 1, 4~($\rho_s$) 
\\
\hline 
\hline
\end{tabular}
\end{center}
\label{tab:parameters}
\end{table}

In PF simulations, a solid seed with $U=0$ is initialized within a liquid domain at $U=-\Omega$.  
The seed is shaped as a quarter of a circle, located at the bottom-left corner of the domain. 
Symmetric (i.e., no-flux) boundary conditions are applied on the four domain boundaries. 

In DNN simulations, the initial solute field is $u=\Omega$, except for $u=0$ on the needle. 
We use $\Delta x = \rho_s$ for the higher $\Omega=0.2$ and $\Delta x = 4 \rho_s$ for the other $\Omega=0.05$ and 0.1. 
Those simulations use symmetric boundary conditions in all directions.


For PF simulations using a solute supersaturation $\Omega=0.05$, the grid spacing is $\Delta x = W = 0.07 \rho_s$ with $W = 400 d_0$, and the radius of the initial solid seed is $100 d_0 \simeq 0.0175 \rho_s$. 
The simulation domain is a channel of width $\widetilde{\Lambda}  =  4.2, 10.9$, or $44.5$ in the $y$ direction and length $2801.0\rho_s$ in the $x$ direction. 
The simulated time is $\tilde{t} = 110$. 
The corresponding DNN simulations are initialized with two needles of length 3$\rho_s$, pointing towards positive $x$ and $y$ directions, and rooted at the bottom left corner. 
They grow until $\tilde{t} = 1100$ within a domain of $N_x \times N_y = 8000\rho_s \times 25 \rho_s$, where $N_x$ and $N_y$ are the domain sizes in $x$ and $y$ directions, respectively. 

Next, in order to investigate the effect of the channel width on dendrite growth dynamics, we use a supersaturation $\Omega = 0.2$ for PF simulations. Those simulations use $\Delta x= W =30d_0$ with $\widetilde{\Lambda}$ varying from 13.1 to 214.8 for a channel of length 1075.46$\rho_s$. 
The seed radius is $100d_0\simeq0.28\rho_s$ and the simulated time is $\tilde{t} = 200$. 
DNN simulations are performed with a single needle located in the center of the channel. 
We explore different channel width of 61, 125, 157, 213, 253, 509, and 4093$\rho_s$ with the length 27069$\rho_s$ for $\Omega = 0.2$. 
For $\Omega = 0.1$, the widths are 192, 308, 628, 756, 884, 1268, and 2548$\rho_s$ and the length of 36088$\rho_s$. 
The widths are 116, 500, 2036, 2804, 3188, 4084, and 8180$\rho_s$ and the length of 40960$\rho_s$ for $\Omega = 0.05$. 
In all simulations, the needle initially has a length $3 \Delta x$, and grows on the $x$ direction until $\tilde{t} =25000$.

\section{Result}
\subsection{Needle growth at $\Omega = 0.05$}

\begin{figure}[!b]
\centering
\includegraphics[width=\columnwidth]{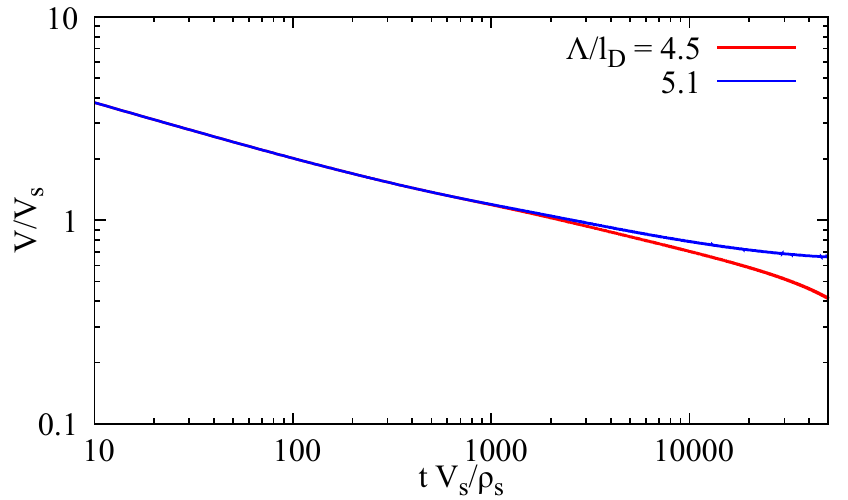}
\caption{ 
Needle growth within ${\Lambda}/l_D = 4.5$ (red line) and 5.1 (blue line) at $\Omega = 0.05$. 
}
\label{fig:long}
\end{figure}

We showed that the needle within a channel shows two growth patterns in Fig. 3 in the main mainuscript. 
When the channel width is lower than the critical spacing $\Lambda < \Lambda_c = 5 l_D$, the needle growth follows $V \sim t^{-2/3}$. 
On the other hand, the needle reaches a steady state for a wide channel. 

For the low $\Omega = 0.05$ in Fig. 3(d), the final growth patterns near $\Lambda_c$ are not clear by $\tilde{t} = 25000$. 
Hence, we perform additional simulations with $\Lambda/\rho_s$ = 2804 and 3188 ($\Lambda/l_D=$ 4.5 and 5.1, respectively) with a longer time $\tilde{t} =50000$. The results are shown in Fig.~\ref{fig:long}. 
In this log-log plot, the needle within ${\Lambda}/l_D = 4.5$ (red line) decelerates continuously, and the deceleration rate increases at the end of the simulation. 
On the other hand, for ${\Lambda}/l_D = 5.1$ (blue line), the needle tends to approach a steady state velocity between 0.6 and 0.7 $V_s$ as we observed in the other simulations. 
Therefore, those needles at $\Omega = 0.05$ agree with the suggested critical spacing $\Lambda_c = 5 l_D$; A dendrite needle within a narrow channel ${\Lambda} < \Lambda_c$ decelerates while the needle approaches a steady state velocity when the channel width is larger than $5 l_D$
